\documentclass{aastex701}

\usepackage{amsmath}
\usepackage{wasysym}

\submitjournal{ApJs}

\begin{document}

\title{Coronal mass ejection arrival forecasting with the drag-based assimilation of satellite observations}

\author[orcid=0009-0009-1264-4202,gname=Zaina, sname='Abu-Shaar']{Zaina Abu-Shaar}
\affiliation{Center for Digital Engineering, Bolshoy Boulevard 30, bld. 1, Moscow 121205, Russia}
\email[show]{z.abu-shaar@rcdei.com}  

\author[orcid=0000-0002-9189-1579,gname=Tatiana, sname='Podladchikova']{Tatiana Podladchikova} 
\affiliation{Center for Digital Engineering, Bolshoy Boulevard 30, bld. 1, Moscow 121205, Russia}
\email[show]{t.podladchikova@rcdei.com}  

\author[orcid=0000-0003-2073-002X,gname=Astrid M., sname='Veronig']{Astrid M. Veronig}
\affiliation{Institute of Physics, University of Graz, Universit{\"a}tsplatz 5, A-8010 Graz, Austria}
\affiliation{Kanzelh{\"o}he Observatory for Solar and Environmental Research, University of Graz, Kanzelh{\"o}he 19, 9521, Treffen, Austria}
\email{astrid.veronig@uni-graz.at}

\author[0000-0002-8680-8267,sname=Mateja,gname=Dumbovi\'c]{Mateja Dumbovi\'c}
\affiliation{Hvar Observatory, Faculty of Geodesy, University of Zagreb, Kaciceva 26, HR-10000 Zagreb, Croatia}
\email{mdumbovic@geof.hr}

\author[orcid=0000-0001-7662-1960,gname=Stefan J., sname='Hofmeister']{Stefan J. Hofmeister}
\affiliation{Columbia University, New York, NY 10027, USA}
\email[show]{sjh2211@columbia.edu}  

\begin{abstract}
Forecasting the arrival of coronal mass ejections (CMEs) is vital for protecting satellites, power systems, and human spaceflight. We present HELIOPANDA: Heliospheric Observer for Predicting CME Arrival via Nonlinear Drag Assimilation, a framework that integrates the Drag-Based Model (DBM) with spacecraft observations using iterative parameter estimation and Kalman filter assimilation. We introduce a method for estimating the solar wind speed $w$ and drag parameter $\gamma$, two key but usually unknown quantities controlling CME propagation, through direct solutions of the DBM equations. We tested the method on 4,480 synthetic CME profiles spanning CME speeds of $200-3500$~km/s, solar wind speeds of $250-800$ km/s, and drag parameters of $0.1-1.0\times10^{-7}$~km$^{-1}$. The results demonstrate that the framework provides accurate reconstructions of the DBM input parameters, providing a solid basis for in-situ and remote-sensing applications. By testing a single virtual spacecraft positioned at nine distances along the Sun–Earth line, HELIOPANDA achieved arrival-time errors as low as 0.6 hours for a 600 km/s CME and 1 hour for a 2500 km/s CME when the spacecraft was located 30 million km from the Sun. We developed a Kalman filter framework to assimilate noisy heliospheric data into the DBM, enabling recursive updates of CME kinematics and robust estimates of $w$ and $\gamma$, and yielding Earth and Mars arrival-time predictions within $1-2$ hours using 160 simulated hourly measurements. By combining DBM, parameter recovery, and data assimilation, HELIOPANDA provides a pathway to real-time, multi-point CME forecasts, suited to observations from Solar Orbiter, Parker Solar Probe, PUNCH, and planned L4/L5 missions.
\end{abstract}


\section{Introduction}
Coronal mass ejections (CMEs) are massive eruptions from the solar corona, ejecting up to $10^{13}$~kg of magnetized plasma, propelled by magnetic forces (see reviews by \citealp{Forbes2006,Vrsnak2008,Chen2011, Webb2012, Cheng2017, Green2018}). Traveling in interplanetary space at speeds of about 100–3500 km/s \citep{Gopalswamy2009,Michalek2009,Tsurutani2014,Veronig2018,RodriguezGome2020}, CMEs are among the most energetic phenomena in the Solar System, releasing energies up to $10^{32}$ erg \citep{Vourlidas2010,Emslie2012}.

Predicting CME arrival times at planets or other targets in the Solar System is a critical component of space weather prediction, with direct implications for satellite operations, power grid resilience, and human spaceflight safety \citep{Schwenn2005,Baker2013,Durante2011}. When directed toward Earth or Mars, CMEs can drive geomagnetic storms at Earth and trigger intense space weather disturbances at Mars, including ionospheric compression, atmospheric escape, and localized auroras.

Forecasting CME propagation and arrival relies on a number of complementary modeling strategies. These methods can broadly be grouped into three categories. First, purely empirical and statistical approaches, often called kinematical-empirical models, estimate CME arrival times by relating coronagraphically measured parameters to interplanetary CME (ICME) travel times and heliospheric evolution (e.g., \citealp{Gopalswamy2000, Vrsnak2002,Manoharan2011}). Second, global three-dimensional magnetohydrodynamic (MHD) simulations, such as ENLIL \citep{Odstrcil2003} and EUHFORIA \citep{Pomoell2018}, capture the heliospheric response to CMEs and the shocks they drive, though they remain computationally demanding and sensitive to uncertain CME initial conditions (e.g., speed, geometry, and launch direction). Complementing these modeling approaches, extensive recent research has focused on reconstructing CME trajectories and characterizing their early evolution, emphasizing the need to estimate CME speed, mass, and propagation direction before they enter interplanetary space \citep{Gopalswamy2000, Bein2011, Veronig2018, Dissauer2019, Chikunova2020, Gou2020, Jain2024_DIRECD, Jain2024_May8, Podladchikova2024_Three_Part,Veronig2025LRSP}. 

Third, bridging the gap between empirical/statistical approaches and global 3D MHD simulations is a class of analytical, kinematical models based on MHD or hydrodynamic considerations. These approaches assume that beyond a certain distance (typically beyond 20$R\astrosun$, \citealp{Sachdeva2015}), CME dynamics are dominated by drag-like interactions with the ambient solar wind. A widely used model is the Drag-Based Model (DBM, \citealp{Vrsnak2013, Vrsnak2014}), which provides analytical solutions for CME distance and speed profiles by incorporating the observed tendency of fast CMEs to decelerate and slow CMEs to accelerate toward the solar wind speed (e.g., \citealp{Gopalswamy2001,Vrsnak2004,Owens2004,Lara2009}). DBM provides analytical solutions for CME distance and speed profiles, enabling estimates of arrival times. However, its predictive accuracy depends heavily on the values of two key parameters: the solar wind speed $w$ and the drag parameter $\gamma$. These parameters are typically unknown in real-time scenarios and are often assumed or manually tuned, introducing significant uncertainty. Average solar wind speeds typically lie in the range of 300 to 800 km/s (e.g., \citealp{Schwenn2005,Richardson2018,Petrukovich2020}), and a variety of physics-based, empirical, and coronal-hole models provide forecasts of the background solar wind \citep{Arge2000,Vrsnak2007,Owens2008, Rotter2015,Reiss2016,Pomoell2018,Hofmeister2020,Nitti2023}. However, these forecasts may be affected by uncertainties in both accuracy and timing, which can translate into several hours of uncertainty in CME arrival predictions. Moreover, in-situ measurements of the solar wind are currently restricted to specific vantage points, such as L1, Parker Solar Probe, Solar Orbiter, or STEREO, but not continuously along the Sun--Earth line, where CME drag acts most strongly.  Consequently, this dual uncertainty in the background speed $w$ and the drag parameter $\gamma$ represents a major limitation for drag-based forecasting \citep{Vrsnak2013,Zic2015}.

In contrast, the CME initial speed near the Sun is typically better constrained through remote sensing observations such as coronagraph images, making it a more reliable input for modeling CME propagation. However, uncertainties in drag-related parameters $w$ and $\gamma$ remain significant challenges for accurate arrival-time forecasting.

Recent advances aim to overcome these limitations through the ensemble DBM approach \citep{Mays2015,Dumbovic2018,Dumbovic2021,
Calogovic2021,Calogovic2025}, which quantifies prediction uncertainty by sampling a range of CME and solar wind parameters. Additionally, data assimilation techniques that merge physics-based solar wind models with in-situ spacecraft observations have enhanced operational forecasts \citep{Lang2019}, while uncertainty analyses of reconstructed CME parameters further constrain model inputs \citep{Verbeke2023}.

The increasing presence of heliophysics missions, including Parker Solar Probe, Solar Orbiter, STEREO, and PUNCH, provides unprecedented opportunities for continuous, multi-point solar wind and CME monitoring. Recent studies also emphasize the importance of L4/L5-based observations for improving such forecasts \citep{Lee2024,Lugaz2025,West2025,Podladchikova2025_L5}. Upcoming and proposed missions, such as ESA’s Vigil at L5, Korea’s L4 mission, and the $4{\pi}$ polar/global perspective concept for tracking solar wind and CMEs, further underscore the growing drive toward continuous, multi-point coverage. Together, these initiatives  provide in-situ and remote-sensing observations across diverse heliocentric distances, supporting operational forecasting and the development of advanced data-assimilative models.

In this study, we present HELIOPANDA: Heliospheric Observer for Predicting CME Arrival via Nonlinear Drag Assimilation, a next-generation framework that combines the physics-based DBM with spacecraft observations to predict CME arrival. Two of the key DBM parameters, the solar wind speed $w$ and the drag coefficient $\gamma$, that determine arrival time are typically unknown and have previously been inferred from CME kinematics via least-squares fitting \citep[e.g.,][]{Zic2015}. To address this challenge, the HELIOPANDA framework advances in three steps. First, we develop and validate a method for estimating $w$ and $\gamma$ from pairs of distance and speed observations through direct solutions of the DBM equations (Section~\ref{Section_w_gamma_est}). Second, as a proof of concept, we apply this method to virtual in-situ probes by treating the case of a single spacecraft placed independently at nine different heliocentric distances along the Sun--Earth line (Section~\ref{Section_in_situ}). Finally, we extend the approach to heliospheric remote-sensing by assimilating simulated noisy imager observations with a Kalman filter, enabling dynamic reconstruction of CME propagation and arrival-time predictions at Earth and Mars (Section~\ref{Section_remote_sensing}).

\section{Direct estimation of solar wind speed and drag parameter from the Drag-Based Model}\label{Section_w_gamma_est}

In this section, we introduce a method for estimating the solar wind speed $w$ and the drag parameter $\gamma$, two key parameters that strongly affect CME propagation, through direct solutions of the DBM equations. To assess its performance, we test whether the method can correctly recover the background values of $w$ and $\gamma$. For this purpose, we generate synthetic CME profiles using the analytical DBM equations, where both parameters are specified. These synthetic runs provide the necessary inputs (CME distance and velocity at successive times) along with the known ``true'' values. We then apply our iterative solver at each time step across the full Sun--Earth interval, starting from a wide range of initial guesses, to evaluate whether the method converges to the correct solution. Because convergence behavior may vary with both time and initial conditions, validating across the entire propagation ensures the robustness of the method. Establishing this reliability under controlled conditions provides the foundation for applying the method to in-situ and noisy remote-sensing spacecraft observations for arrival-time forecasting. Here we treat CME arrival as the arrival of its front rather than the CME-driven shock, consistent with the DBM formulation.

The DBM assumes that, beyond the solar corona, CME dynamics are dominated by MHD drag — a collisionless transfer of momentum and energy between the CME and the solar wind via MHD waves, which takes the same quadratic form as aerodynamic drag: fast CMEs decelerate, slow ones accelerate \citep{Vrsnak2013,Zic2015}, and the acceleration follows a quadratic form:
\begin{equation}\label{eq_motion_a}
a = -\gamma (v - w)|v - w|.
\end{equation}
Here, $v$ is the CME speed, $w$ is the solar wind speed, and $\gamma$ is the drag parameter, which depends on the CME mass, size, and solar wind density, and is given by:
\begin{equation}\label{eq_drag_gamma}
\gamma=\frac{c_{d}A\rho_{w}}{M+M_v}.
\end{equation}
Here, $c_{d}$ is the dimensionless drag coefficient, $A$ is the CME cross-sectional area, $\rho_{w}$ is the ambient solar-wind density, $M$ is the CME mass, and $M_{v} \sim p_{w}V/2$ is the so-called virtual mass associated with the CME volume $V$ \citep{Vrsnak2013}.   

Assuming constant $w$ and $\gamma$, which is valid mainly beyond 20$R\astrosun$, the CME kinematics can be obtained by basic integration of Eq. \eqref{eq_motion_a}, resulting in the following relations:

\begin{equation}\label{eq_DBM_r}
r(t) = \pm \frac{1}{\gamma} \ln \left[1 \pm \gamma (v_0 - w)t\right] + w t + r_0,
\end{equation}

\begin{equation}\label{eq_DBM_v}
v(t) = \frac{v_0 - w}{1 \pm \gamma (v_0 - w) t} + w.
\end{equation}
Here, $r_{0}$ and $v_{0}$ are the CME position and speed, respectively, at the start of the drag-dominated phase. The plus sign applies for decelerating CMEs with $v_{0}>w$ and the minus sign applies for accelerating CMEs with $v_{0}<w$.

Due to the non-linear nature of these equations, an explicit solution for $w$ and $\gamma$ is not analytically available. We therefore linearize both expressions using a first-order Taylor expansion around an initial guess denoted by $(w_0, \gamma_0)$:
\begin{equation}
r(t) - r_0 \approx \frac{\partial r}{\partial w} \bigg|_{w_0, \gamma_0} (w - w_0) + \frac{\partial r}{\partial \gamma} \bigg|_{w_0, \gamma_0} (\gamma - \gamma_0),
\end{equation}
\begin{equation}
v(t) - v_0 \approx \frac{\partial v}{\partial w} \bigg|_{w_0, \gamma_0}(w - w_0) + \frac{\partial v}{\partial \gamma} \bigg|_{w_0, \gamma_0} (\gamma - \gamma_0).
\end{equation}
These equations can be written compactly as a system of linear equations in matrix form:
\begin{equation}\label{eq_linear_system}
\left[
\begin{array}{cc}
\frac{\partial r}{\partial w} & \frac{\partial r}{\partial \gamma} \\
\frac{\partial v}{\partial w} & \frac{\partial v}{\partial \gamma}
\end{array}
\right]
\left[
\begin{array}{c}
w - w_0 \\
\gamma - \gamma_0
\end{array}
\right]
=
\left[
\begin{array}{c}
r(t) - r_0 \\
v(t) - v_0
\end{array}
\right].
\end{equation}
Solving this system yields:
\begin{equation}\label{w_est}
w = \frac{ \frac{\partial v}{\partial \gamma} (r(t) - r_0) - \frac{\partial r}{\partial \gamma} (v(t) - v_0)}{\frac{\partial r}{\partial w} \cdot \frac{\partial v}{\partial \gamma}
-
\frac{\partial v}{\partial w} \cdot \frac{\partial r}{\partial \gamma}} + w_0,
\end{equation}
\begin{equation}\label{gamma_est}
\gamma = \frac{-\frac{\partial v}{\partial w} (r(t) - r_0) + \frac{\partial r}{\partial w} (v(t) - v_0)}{\frac{\partial r}{\partial w} \cdot \frac{\partial v}{\partial \gamma}
-
\frac{\partial v}{\partial w} \cdot \frac{\partial r}{\partial \gamma}} + \gamma_0.
\end{equation}

\begin{figure*}[ht!]
\centering
\includegraphics[width=0.7\textwidth]{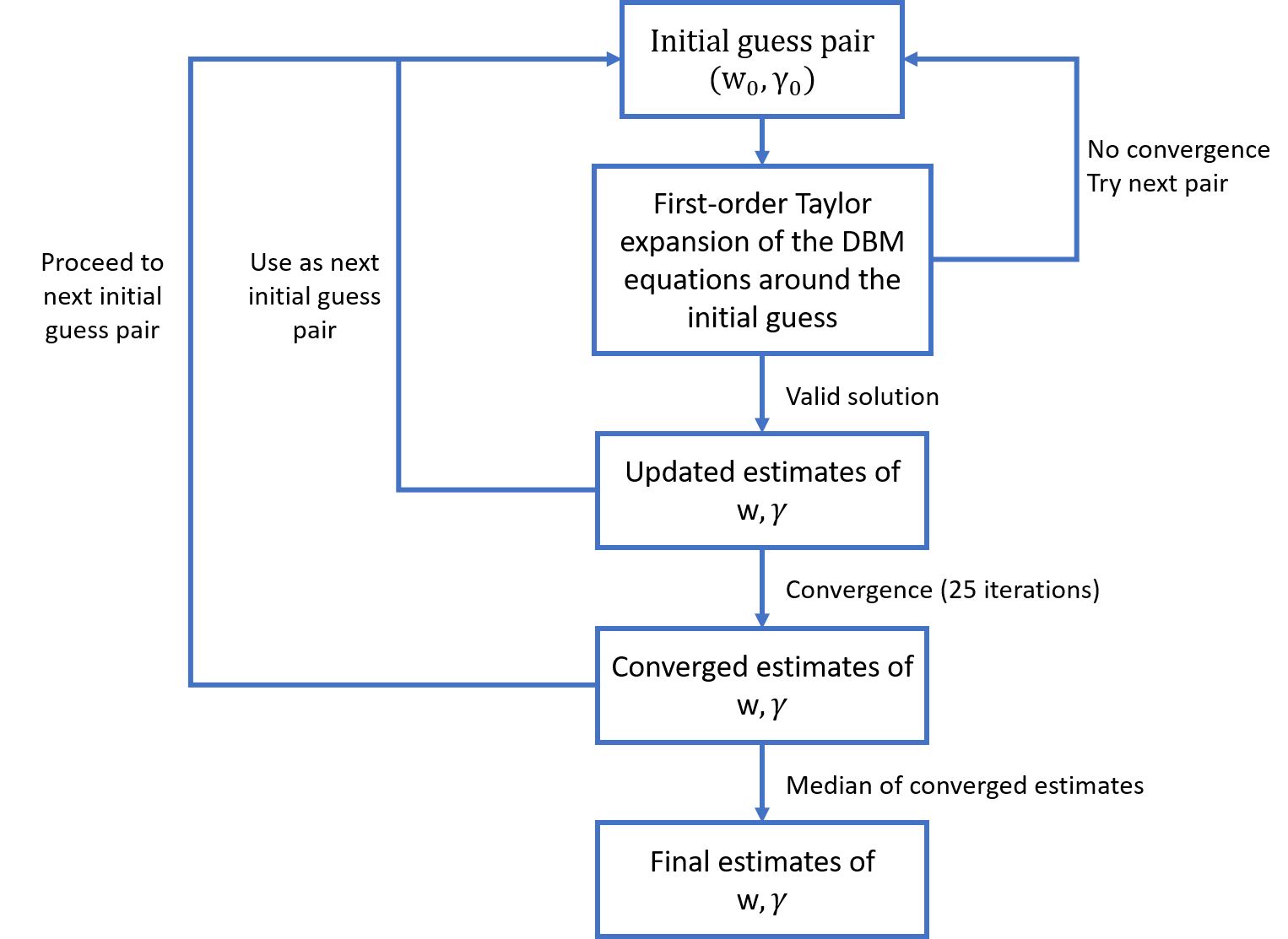}
\caption{Block diagram illustrating the iterative procedure used to estimate the solar wind speed $w$ and the drag parameter $\gamma$ from the DBM equations.
\label{fig:block_diagram_estimates_solution}}
\end{figure*}

This solution for the linearized system provides estimates of $w$ and $\gamma$ for a given initial guess $(w_0, \gamma_0)$. However, the nonlinear nature of Equations~\eqref{eq_DBM_r} and~\eqref{eq_DBM_v} necessitates an iterative solution method. We employ an iterative update approach, illustrated schematically in Figure~\ref{fig:block_diagram_estimates_solution}. The process begins with an initial guess for $w$ and $\gamma$. Using this initial pair,  \((w_0, \gamma_0)\), we linearize the equations via a first-order Taylor expansion as described above and subsequently solve the resulting linear system to obtain updated estimates of $w$ and $\gamma$. These updated estimates are then used as the initial guess for the next iteration. This iterative update continues sequentially for a total of 25 iterations, or until a stopping condition is triggered. Stopping conditions include: the argument of the logarithmic term in Eq.~\eqref{eq_DBM_r} becomes zero or negative, which renders the expression undefined; division by zero occurs in Eq.~\eqref{w_est} or Eq.~\eqref{gamma_est}; or the estimated parameters fall outside physically reasonable bounds. Specifically, we constrain the solar wind speed $w$ to lie between 250 and 800~km/s, and the drag parameter $\gamma$ to lie within the range from \( 0.1 \times 10^{-7} \) to \( 1.0 \times 10^{-7} \)~km\(^{-1} \)~\citep{Vrsnak2013}. 

For each initial guess pair $(w_0, \gamma_0)$, we apply the iterative estimation procedure independently. If a given pair does not converge after 25 iterations, we flag it as non-convergent, but we still test the other pairs. This ensures that all initial guesses in the grid are evaluated for convergence. Once we achieve convergence for a given pair, we store the resulting estimates. After we evaluate all initial guess pairs, we compute the final values
of $w$ and $\gamma$ as the median of all the converged estimates. The complete code is publicly available \citep{astrozee_2025}.

To evaluate the convergence behavior of the proposed method, we simulated a wide set of CME propagation scenarios using the DBM equations, generated synthetic distance and speed measurements at multiple points along their paths, and applied our inversion procedure to recover the background solar wind speed $w$ and drag parameter $\gamma$ for a broad range of parameter sets ($v_{0}$, $w$, $\gamma$). We considered a total of 4,480 CME profiles, each generated using different combinations of $v_{0}$, $w$, and $\gamma$. The CME initial speeds were varied over a range from 200 to 3500~km/s, while the solar wind speeds $w$ were varied over a range from 250 to 800~km/s, and the drag parameter $\gamma$ was varied over a range from \( 0.1 \times 10^{-7} \) to \( 1.0 \times 10^{-7} \)~km\(^{-1} \). Each profile corresponds to a time series of 120 observations, sampled at 1-hour intervals.

For each CME profile, we applied the iterative estimation procedure by performing a grid search over initial guesses of $w$ and $\gamma$. We sampled initial guesses for $w$ from 250 to 1000~km/s in 10~km/s increments, and varied initial guesses for $\gamma$ from \( 0.1 \times 10^{-7} \)~km\(^{-1} \) to \( 1.0 \times 10^{-7} \)~km\(^{-1} \), with increments of \( 0.05 \times 10^{-7} \)~km\(^{-1} \). We extended the initial guess range for $w$ to 1000~km/s beyond typical solar wind speeds (800~km/s) to ensure broader convergence coverage and avoid missing solutions near the upper physical limit. This setup enabled us to assess the sensitivity of the solution to initial guesses and to quantify the convergence rate across a broad parameter space.

\begin{figure*}[ht!]
\centering
\includegraphics[width=0.7\textwidth]{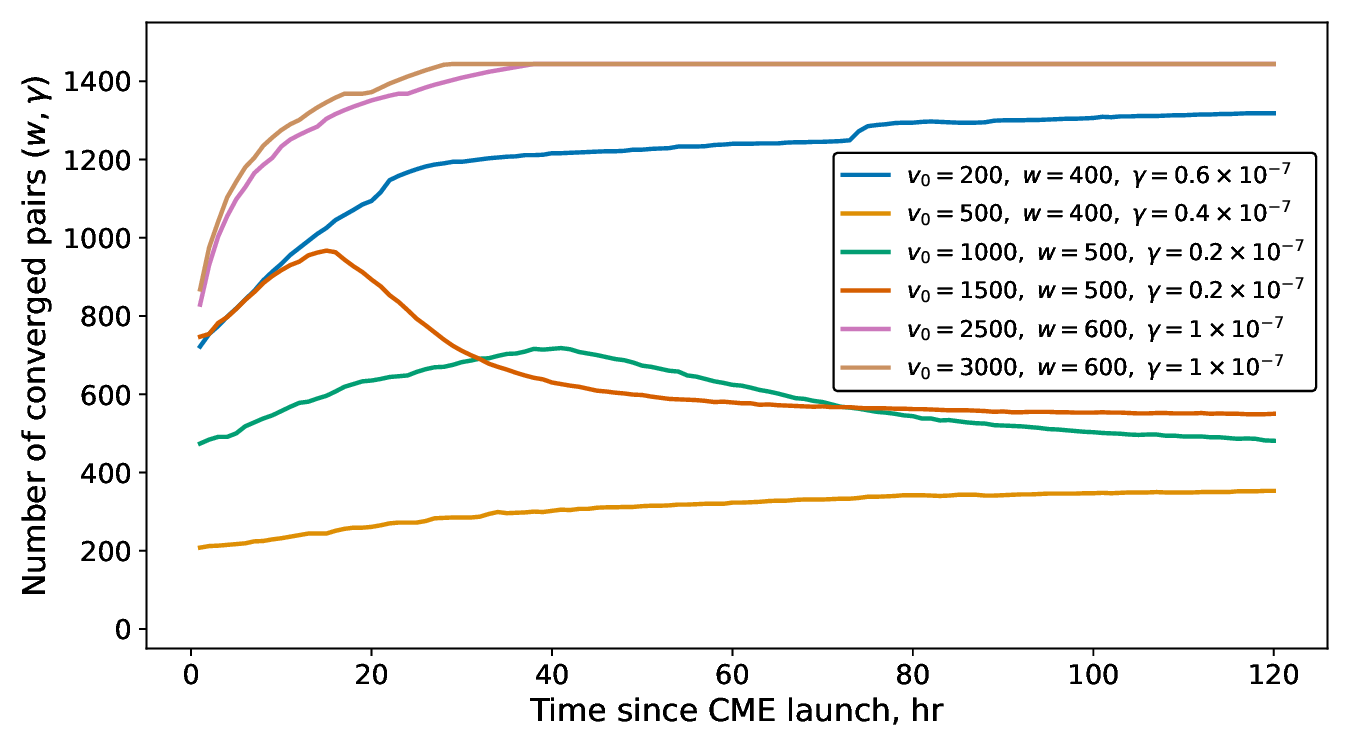}
\caption{Convergence performance of the iterative estimation method for six representative CME parameter cases. The Y-axis shows the number of successful convergences out of 1,444 tested initial guess pairs at each time step (X-axis). Importantly, across all tested combinations of $v_{0}$, $w$, and $\gamma$, there are always initial guesses that achieve convergence at every time step.
\label{fig:counts}}
\end{figure*}

Figure~\ref{fig:counts} illustrates the convergence behavior of the iterative estimation method for six representative cases of CME parameters. For each time step, a comprehensive grid of 1,444 initial condition pairs was tested, and the figure shows the number of these pairs that successfully converged to a solution. Although one might expect the number of converged initial guess pairs to increase and then saturate with time, in some cases it dips or fluctuates. This can happen when Equations~(\ref{eq_linear_system}) become temporarily ill-conditioned, for example, when the factor $\gamma(v_{0}-w)t$ is very small or very large, making its derivatives nearly proportional. Another situation that can reduce convergence is when the CME speed $v$ becomes close to the solar wind speed $w$; in this regime, changes in $\gamma$ have little effect on $r(t)$ and $v(t)$, so some initial guesses fail to refine to the true values. In addition, certain $(w_{0}, \gamma_{0})$ guesses may violate the sign or positivity requirements of Equations~(\ref{eq_DBM_r}) and (\ref{eq_DBM_v}), or lead to updates outside the physically allowed bounds, causing the iteration to stop. Importantly, across all tested combinations of $v_{0}$, $w$, and $\gamma$, there are always initial guesses that lead to convergence at every time step. This shows that the convergence corresponds to consistent parameter estimates since unstable or non-convergent iterations are naturally excluded.
 
\begin{figure*}[ht!]
\plotone{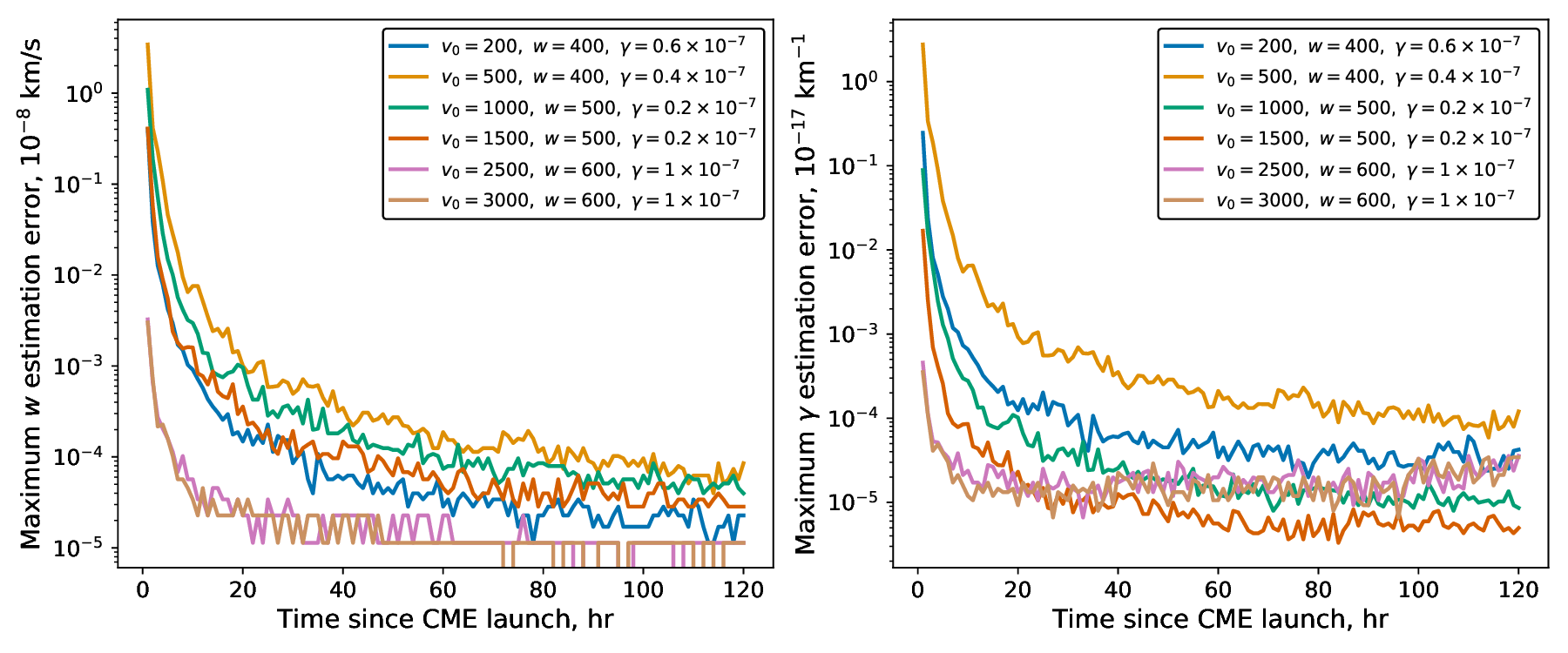}
\caption{Maximum estimation errors of solar wind speed $w$ (left) and drag parameter $\gamma$ (right) over time for six representative CME cases. The y-axes are shown in logarithmic scale. 
\label{fig:max_error}}
\end{figure*}

Figure~\ref{fig:max_error} presents the maximum estimation errors for $w$ and $\gamma$ over the simulation period for the same six representative CME cases shown in Figure~\ref{fig:counts}. The y-axes are plotted in logarithmic scale. For each time step, the maximum absolute error (computed as the difference between the estimated and the true values of $w$ and $\gamma$) is taken among all initial guess pairs that successfully converged. 
As illustrated in Figure~\ref{fig:max_error}, the estimation errors associated with these converged solutions remain extremely low. Across all the 4,480 tested cases, the maximum estimation error of $w$ over the entire propagation period is only $0.0002~\mathrm{km/s}$, and that for $\gamma$ is $0.000004 \times 10^{-7}$~km$^{-1}$ excluding the case when $v_{0} = w$. For this case, the estimation error of $\gamma$ could increase to 0.012 km$^{-1}$ as the derivatives in Equation(\ref{eq_linear_system}) may become proportional, rendering the system nearly singular and reducing the precision of parameter estimates.
At early times, the errors are slightly larger near the Sun. In this region, the factor $\gamma(v_{0}-w)t$ in Equations~ (\ref{eq_DBM_r}) and (\ref{eq_DBM_v}) is small, so $r(t)$ and $v(t)$ 
evolve almost linearly with time. When this happens, the derivatives in Equations~(\ref{eq_linear_system}) become nearly proportional, meaning that the distance and speed equations carry almost the same information about $w$ and $\gamma$. As a result, the system provides less independent information to separate the two parameters, and their estimates become less precise.

The proposed iterative scheme described here is particularly well suited for the problem of estimating $w$ and $\gamma$ because it leverages its analytical structure. Using closed-form relations and exact derivatives for $r$ and $v$, combined with Newton-type updates over a scanned range of initial values, the method avoids the sensitivity to starting points typical of Levenberg–Marquardt and the high computational overhead of genetic algorithms. Therefore, this approach ensures a stable convergence within feasible parameter bounds, even when initial estimates are uncertain, providing a solid foundation for applying the method to in-situ and noisy remote-sensing spacecraft observations.

\section{Coronal mass ejection arrival forecasting through virtual in-situ probes along the Sun--Earth line}\label{Section_in_situ}

\begin{figure*}[ht!]
\plotone{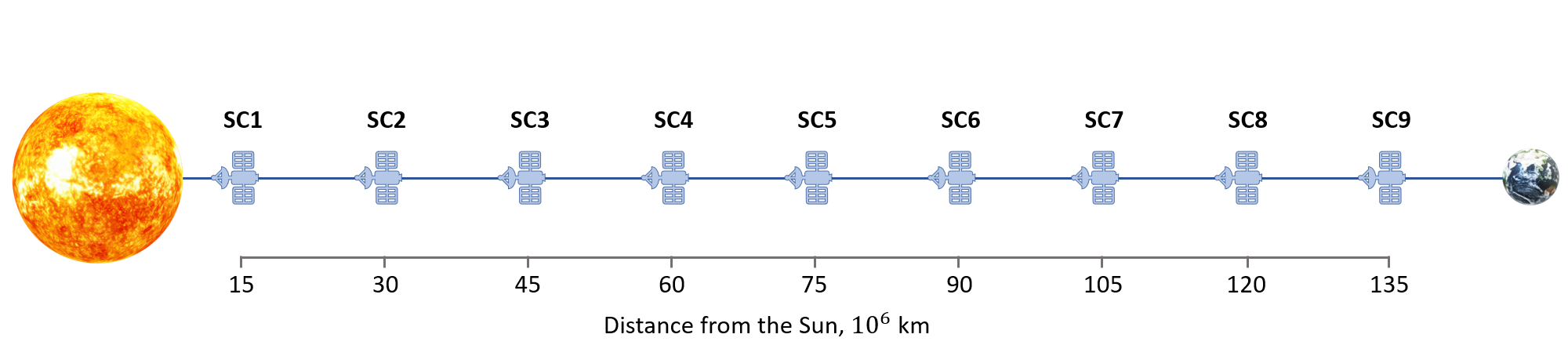}
\caption{Schematic representation of in-situ probes positioned along the Sun--Earth line to observe Earth-directed CMEs.
\label{fig:SC_sketch}}
\end{figure*}

At present, measurements of CME distance, velocity, and the background solar wind speed are available mainly near Earth (L1, STEREO) or at limited heliocentric distances (e.g., Parker Solar Probe, Solar Orbiter). While continuous coverage along the Sun–Earth line is challenging, even a single spacecraft positioned between the Sun and Earth could provide valuable information. To assess the potential impact of such observations, we simulate a scenario in which nine spacecraft are positioned equidistantly along the Sun--Earth line to measure the properties of Earth-directed CMEs (Figure~\ref{fig:SC_sketch}). When a spacecraft encounters a CME, it provides in-situ measurements of the CME’s propagation speed and its heliocentric distance, the latter determined directly from the spacecraft’s position. Each spacecraft position is treated independently, so the results can be interpreted as those from a single spacecraft placed at different distances from the Sun rather than from a simultaneous multi-spacecraft observation. We assume both distance and speed include a small, controlled level of noise to account for instrumental and observational uncertainties.
For the simulation, we model the measurement noise as Gaussian with a standard deviation of $\sigma_r$ = 1000 km for the distance measurements and $\sigma_v$ = 5 km/s for the velocity measurements, representing instrumental and fitting uncertainty. The $\sigma_r$ value is chosen to exceed orbit-determination accuracies: $\sim$$2-10$ km for Lagrange-point missions, with Delta Differenced One-Way Range (DDOR) improving position uncertainty by $>25\%$ \citep{Beckman2003orbit}. STEREO achieved orbit determination accuracies of about $1-3$ km in the first 24 hours, $0.24-0.32$ km during Earth-centered phasing loops, and $1.2-2.75$ km in heliocentric cruise, with $30$-day predictions typically $<190 - 340$~km \citep{Mesarch2007orbit}. For Parker Solar Probe, the 1,200 km ($3\sigma$) value is the target orbit determination uncertainty within 0.25~AU \citep{Drew_Orbit2018}. Regarding velocity measurements and $\sigma_v$, calibration of the Wind/SWE Faraday Cup indicates velocity uncertainties $\lesssim 0.16\%$ in magnitude \citep{Kasper2006}. Validation of DSCOVR’s Faraday Cup shows $\sim4\%$  accuracy for the radial $V_{x}$ component, which dominates the solar wind speed magnitude, at typical solar wind speeds, with $r\approx0.96$ compared to Wind and ACE \citep{Lotoaniu2022}. Cross-comparison of ACE real-time and OMNI final data during disturbed periods further shows that, for 97\% of the cases, most speed deviations are only a few km/s \citep{Podladchikova2018_StormFocus}. Thus, $\sigma_v$ = 5~km/s safely over-bounds instrumental errors across multiple spacecraft. Continued advances in in-situ plasma instrumentation and multi-point observations will further reduce these uncertainties and improve solar wind characterization. We acknowledge that instantaneous speed can reflect distinct CME features such as the outer shock, leading edge, and inner bright rim/cavity boundary \citep{Veronig2018}, and therefore, when imaging is available, we suggest determining the speeds of these features separately to capture CME structural variability. In the absence of imaging, partial separation of feature speeds may still be inferred from in-situ measurements (e.g., shock, sheath, and ejecta flows), though such associations require modeling support.

We then use these data to estimate the background solar wind speed $w$ and the drag parameter $\gamma$ at each spacecraft position using the iterative method described in the previous section. If a valid solution cannot be found directly from the satellite measurements, it may indicate that the noise in the data prevents convergence. Since initial conditions typically lead to a solution for accurate data, the algorithm then searches within $\pm 30~\mathrm{km/s}$ around the measured velocity $v$ to estimate the DBM parameters.
Among all the solutions found in this vicinity, the one that best matches the measured $r$ and $v$ is selected. The estimated parameters are then applied within Equation~(\ref{eq_DBM_r}) to propagate the CME to 1~AU and predict its arrival time at Earth, with the estimates updated at each measurement point.

To evaluate the performance of this approach, we tested it across a set of CME scenarios that cover a range of initial CME speeds (from slow to fast). Here, we present two representative examples: a CME with an initial velocity of $v_{0}=600$~km/s, propagating through a solar wind with a speed of $w=400$~km/s and a drag parameter of $\gamma = 0.2 \times 10^{-7}$~km$^{-1}$, and a faster CME with $v_{0}=2500$~km/s, propagating through a solar wind of $w=300$~km/s and $\gamma = 0.1 \times 10^{-7}$~km$^{-1}$. In all scenarios, 
$w$ and $\gamma$ are kept constant along the propagation path, consistent with the formulation of the DBM equations used to compute 
$r(t)$ and $v(t)$.

\begin{figure*}[ht!]
\plotone{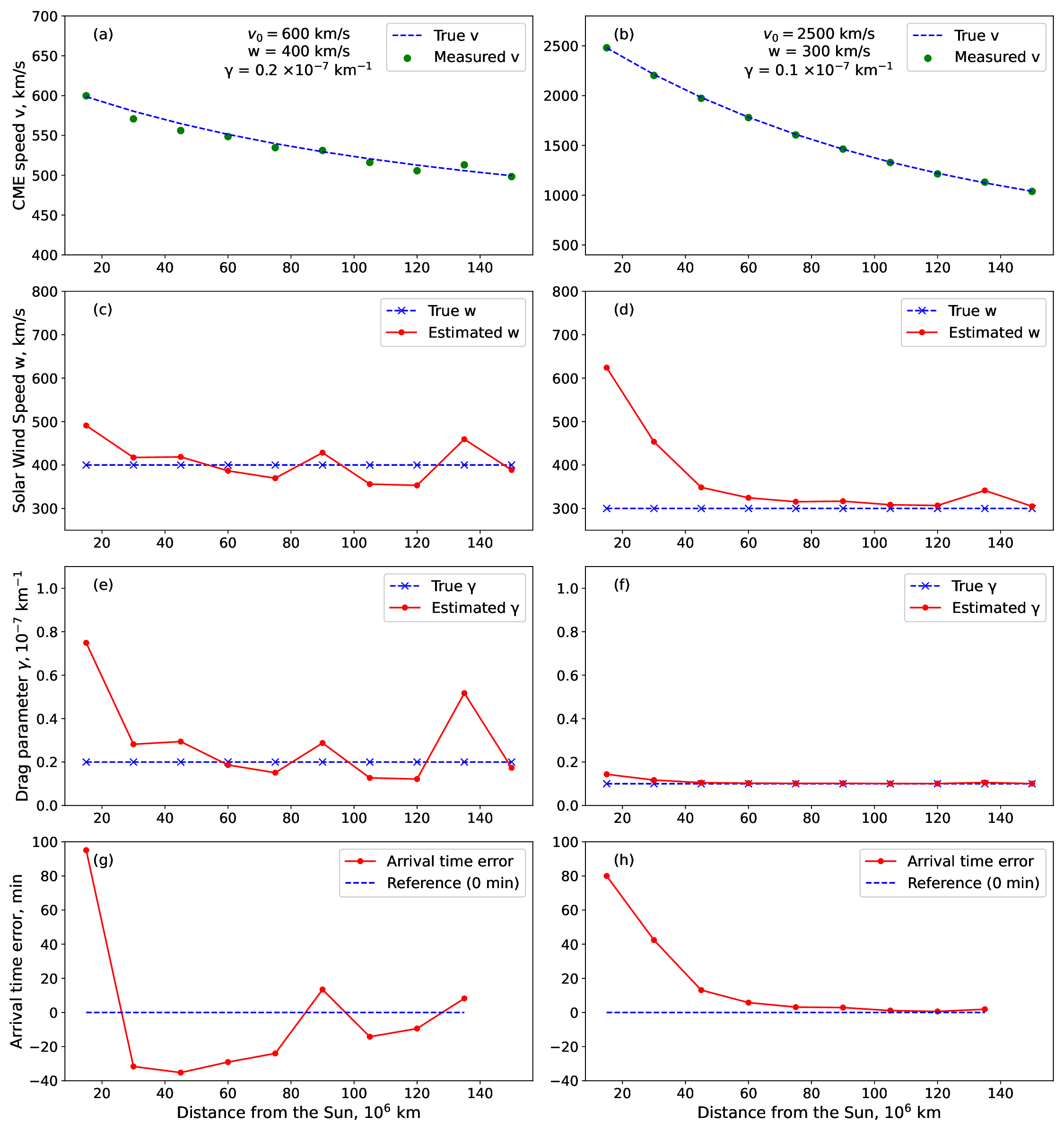}
\caption{Estimation results for two representative CME scenarios observed by spacecraft positioned along the Sun–Earth line. Left panels (a), (c), (e), and (g) correspond to a CME with initial speed 600~km/s, solar wind speed 400~km/s, and drag parameter $\gamma = 0.2 \times 10^{-7}$~km$^{-1}$. Right panels (b), (d), (f), and (h) correspond to a CME with initial speed 2500~km/s, solar wind speed 300~km/s, and drag parameter $\gamma = 0.1 \times 10^{-7}$~km$^{-1}$. Panels (a) and (b) show the CME speed as a function of heliocentric distance and CME speed measurements. Panels (c) and (d) show the estimated solar wind speed $w$ at each spacecraft location. Panels (e) and (f) show the estimated drag parameter $\gamma$. Panels (g) and (h) show the arrival-time estimation error (difference between predicted and actual arrival times in minutes). 
\label{fig:sc_case_ex}}
\end{figure*}

Figure~\ref{fig:sc_case_ex} summarizes the results of $w$ and $\gamma$ estimation and arrival-time prediction at each of the nine spacecraft positions along the Sun–Earth line for the two scenarios. The left panels correspond to the slower CME scenario, while the right panels show results for the faster CME scenario. Panels (a) and (b) display the CME speed as a function of heliocentric distance (dashed blue) for each scenario along with the CME speed measurements obtained by each spacecraft (green dots). Panels (c) and (d) show the estimated values of the solar wind speed $w$, while panels (e) and (f) show the corresponding estimates of the drag parameter $\gamma$ at each spacecraft location. For both scenarios, both parameters are in close agreement with the true values throughout the CME’s propagation. Panels (g) and (h) present the CME arrival-time prediction error at Earth, measured as the difference between the estimated and actual arrival times in minutes. In the fast CME case (panel (d)), the large early errors in $w$ mainly reflect the influence of measurement noise at a stage when the CME motion is still dominated by $v_{0}$ and the effect of $w$ on the trajectory is weak. Nevertheless, in panel (h), the arrival-time estimates remain relatively accurate even from the first data point because the initial $r,v$ measurement, combined with the known $r_{0}$ and $v_{0}$, already constrains the DBM solution sufficiently, and for fast CMEs the transit time is less sensitive to $w$ errors. In this case, $\gamma$ is estimated with high accuracy from the start (panel (f)), further supporting reliable arrival-time forecasts.
Remarkably, the arrival-time estimates are already reasonably accurate from the first spacecraft
positioned close to the Sun because we combine the known initial CME position and speed $(r_{0}, v_{0})$ near the Sun with that spacecraft's $(r,v)$ measurement to solve the DBM for $w$ and $\gamma$. The prediction error then decreases as the CME is tracked farther along the Sun--Earth line, demonstrating the robustness of the method even in the early stages of propagation.

To further assess the stability and reliability of the method, we perform a statistical analysis based on 100 independent simulations for each of the scenarios tested. In each experiment, independent Gaussian noise is added to the synthetic measurements of CME heliocentric distance and propagation speed at each spacecraft position. Then, for each trial and spacecraft location, the iterative method was applied to estimate the solar wind speed $w$, drag parameter $\gamma$, and CME arrival time at Earth.

\begin{figure*}[ht!]
\plotone{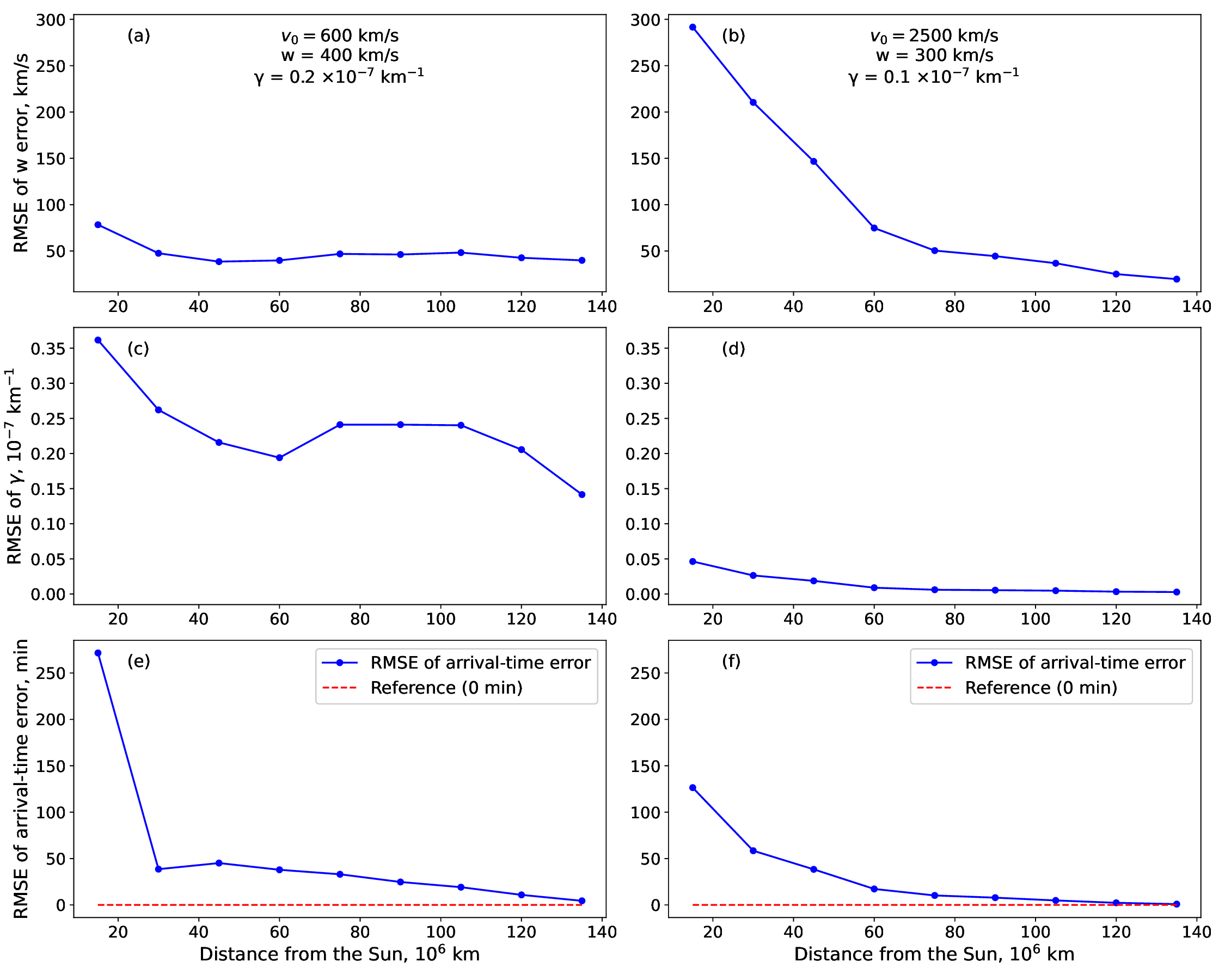}
\caption{Statistical estimation results for two representative CME scenarios observed by spacecraft positioned along the Sun–Earth line. Left panels (a), (c), and (e) correspond to a CME with initial speed 600~km/s, solar wind speed 400~km/s, and drag parameter $\gamma = 0.2 \times 10^{-7}$~km$^{-1}$. Right panels (b), (d), and (f) correspond to a CME with initial speed 2500~km/s, solar wind speed 300~km/s, and drag parameter $\gamma = 0.1 \times 10^{-7}$~km$^{-1}$. Panels (a) and (b) show the RMSEs of the estimated solar wind speed $w$ across the spacecraft array. Panels (c) and (d) show the corresponding RMSEs of the estimated drag parameter $\gamma$. Panels (e) and (f) display the RMSEs of the CME arrival time error. 
\label{fig:sc_stats}}
\end{figure*}

For each simulated position $i$ along the Sun--Earth line, and for each parameter $p$ (representing in turn $w$, $\gamma$, or the arrival time), the error $\varepsilon_{i}^{n}$ in run $n$ is computed as
\begin{equation}
\varepsilon_{i}^{n} = \left(p_{i} - \hat{p}_{i}^{(n)}\right)^{2},
\end{equation}
where $p_{i}$ is the true value and $\hat{p}_{i}^{(n)}$ is the corresponding estimate obtained at the $i$-th position in the $n$-th run. The final reported error is the root-mean-square error (RMSE) over all $N=100$ runs at position $i$:
\begin{equation}
\text{RMSE}_{i} = \sqrt{\frac{1}{N-1} \sum_{n=1}^{N} \varepsilon_{i}^{n}}.
\end{equation}
We applied this general formulation separately to compute the RMSEs for the solar wind speed $w$, the drag parameter $\gamma$, and the CME arrival time. Figure~\ref{fig:sc_stats} presents the results of this statistical evaluation for each of the two representative CME scenarios introduced previously. The left column panels (a, c, and e) correspond to the slower CME case with initial speed 600~km/s, solar wind speed 400~km/s, and drag parameter $0.2 \times 10^{-7}$~km$^{-1}$, while the right column panels (b, d, and f) correspond to the faster CME case with initial speed 2500~km/s, solar wind speed 300~km/s, and drag parameter $0.1 \times 10^{-7}$~km$^{-1}$. Panels (a) and (b) show the RMSEs of the estimated solar wind speed $w$ across the spacecraft array (19.64 to 291.56~km/s). Panels (c) and (d) display the RMSEs of the estimated drag parameter $\gamma$ ($0.0028\times 10^{-7}$ to $0.36\times 10^{-7}$~km$^{-1}$). Panels (e) and (f) depict the RMSEs of the CME arrival-time error in minutes.

The results indicate that the estimation errors for both $w$ and $\gamma$ remain relatively small across all spacecraft positions, with improved accuracy as the CME moves farther from the Sun. Importantly, the arrival-time prediction errors decrease over time. For example, based on the full set of statistical runs shown in Figure~\ref{fig:sc_stats}, the error in the first CME scenario decreased from 4.5 hours at the first spacecraft (15 million km from the Sun) to 0.6 hours at the second (30 million km), while in the second CME scenario it dropped from 2 hours to 1 hour. This demonstrates the approach’s effectiveness even in the early stages of CME propagation. These findings show that even a single spacecraft positioned between the Sun and Earth can provide significantly more accurate CME arrival-time predictions. Since in our analysis each of the nine spacecraft measurements is treated independently, the result for a single spacecraft follows directly from the same dataset without requiring a separate experiment. Because arrival-time prediction errors drop rapidly even with measurements from only one or two spacecraft, missions such as Parker Solar Probe and Solar Orbiter could already provide valuable data for early-stage CME forecasts \citep{Laker2024}. We also emphasize that continuous CME and solar wind measurements along the Sun–Earth line, while challenging, would be highly valuable for future missions.  

\section{Coronal mass ejection arrival forecasting through remote sensing heliospheric measurements}\label{Section_remote_sensing}

In practice, forecasting CME arrival from spacecraft data must account for the current limited availability of in-situ measurements. Missions such as Solar Orbiter and Parker Solar Probe already provide highly valuable observations of CME properties at selected heliocentric distances, but continuous coverage along the entire Sun--Earth line is not feasible with current or near-term missions. Compared to this partial coverage, heliospheric imagers can provide continuous measurements of CME distance in interplanetary space, though these reconstructions are inherently noisy due to projection effects and line-of-sight integration. To make effective use of such data, we introduce a Kalman filter framework. Its role is to assimilate uncertain remote-sensing observations into the DBM, enabling recursive updates of CME kinematics and robust estimation of the background solar wind speed $w$ and drag parameter $\gamma$. Simply injecting noisy HI measurements into the iterative solver can lead to convergence failures or unstable parameter recovery, whereas the Kalman filter mitigates these issues by optimally combining model dynamics with observational uncertainties.

In this section, we simulate a scenario involving 160 remote sensing CME measurements at a cadence of one hour, tracking its heliocentric distance $r$ and velocity $v$ along the Sun--Earth and Sun-Mars trajectories. These simulated observations are representative of data that could be acquired primarily from heliospheric imagers, which are subject to significant uncertainties/noise due to projection effects \citep{Rollett2012,Braga2020}.

To determine meaningful physical parameters such as the solar wind speed $w$ and the drag coefficient $\gamma$ from noisy measurements, we employ a Kalman filter framework. 
The Kalman filter requires that the dynamic system under investigation be described by a mathematical model expressed in state-space form. This representation must accurately capture the evolution of the system states. However, the non-linearity of the original DBM equations complicates direct state-space filtering because the state transition depends nonlinearly on both the CME speed and the unknown parameters $w$ and 
$\gamma$. To enable efficient propagation estimation, we approximate the dynamics of DBM using a linearized motion model that retains the essential physics, particularly the quadratic drag acceleration $a$, while allowing for recursive state updates and noise-resistant parameter estimation. For this approximation, we use the following equations:
\begin{equation}\label{eq_DBM_approx}
  \begin{aligned}
    r_i &= r_{i-1} + v_{i-1}t + \frac{1}{2}a_{i-1}t^2 \\
    v_i &= v_{i-1} + a_{i-1}t.
  \end{aligned}
\end{equation}
Here, acceleration $a$ is defined at each time step $i$ according to Equation~(\ref{eq_motion_a}) and $t$ is the time between step $i$ and step $i-1$.

We present the state-space model for the CME propagation as follows:
\begin{equation}\label{state_eq}
X_i = \Phi X_{i-1} + Ga_{i-1} + \epsilon_i.
\end{equation}
Here, $X_i=[r_{i},v_{i}]^T$ is the state vector at time step $i$, consisting of position $r_{i}$ and velocity $v_{i}$; $\Phi=\begin{bmatrix}
1 & t \\
0 & 1
\end{bmatrix}$ is the transition state matrix that relates the state vector $X_{i-1}$ to the one-step-ahead predicted state vector $X_{i}$; $G=[t^2/2, t]^T$
is the input matrix, which describes the effect of acceleration on the state vector; and $\epsilon_i$ is the uncorrelated and unbiased state noise, accounting for
the imperfect description of CME propagation dynamics by the model, including discrepancies between the linearized approximation (Equation~(\ref{eq_DBM_approx})) and the full non-linear DBM (Equations~(\ref{eq_DBM_r}) and~(\ref{eq_DBM_v})).

To relate the available measurements of the heliospheric distance $r$ and $v$ to the state vector, we introduce the following measurement equation:
\begin{equation}\label{meas_eq}
z_{i} = HX_{i} + \eta_i.
\end{equation}
Here, $z_i=[r_{i}^{m},v_{i}^{m}]^T$ is the measurement vector, consisting of the measured position $r_{i}^{m}$ and the velocity $v_{i}^{m}$; $\eta_i$ is an uncorrelated and unbiased measurement noise, and $H$ is the $2\times2$ identity matrix mapping the state vector $X_i$ to the measurement vector $z_i$.

The complete estimation procedure is illustrated in Figure~\ref{fig:kalman_block}, which summarizes the sequential steps of the proposed DBM-Kalman algorithm and parameter estimation framework. The process begins with initial state estimates for $X_{0}=[r_{0},v_{0}]^T$ and its covariance matrix $P_{0}$, where the diagonal variances for $r_{0}$ and $v_{0}$ are set to 10\% of their true values to reflect initial-state uncertainty. The initial acceleration is set to $a_{0}=0\ \rm{m/s^2}$. At each time step, the algorithm predicts the future state vector,
then assimilates new measurements of the CME’s heliocentric distance $r$ and velocity $v$ to update the estimate of $r$ and $v$ (see Appendix~\ref{appendix_Kalman} for details). From the filtered state, the solar wind speed $w$ and drag parameter $\gamma$ are then derived using the method described in Section~\ref{Section_w_gamma_est}.  If no valid solution can be obtained from the filtered $r$ and $v$, regularization is applied as described below. The difficulty arises because, when the CME velocity $v$ is relatively close to the background solar wind speed $w$, the system of Equations (\ref{eq_linear_system}) becomes nearly singular, so even small estimation errors in 
$r$ and $v$ can lead to large uncertainties in estimating  
$w$ and $\gamma$. To stabilize the estimation of $w$ and $\gamma$, we introduce small regularization terms into Equations~(\ref{w_est})~and~(\ref{gamma_est}):

\begin{figure*}[!htbp]
\plotone{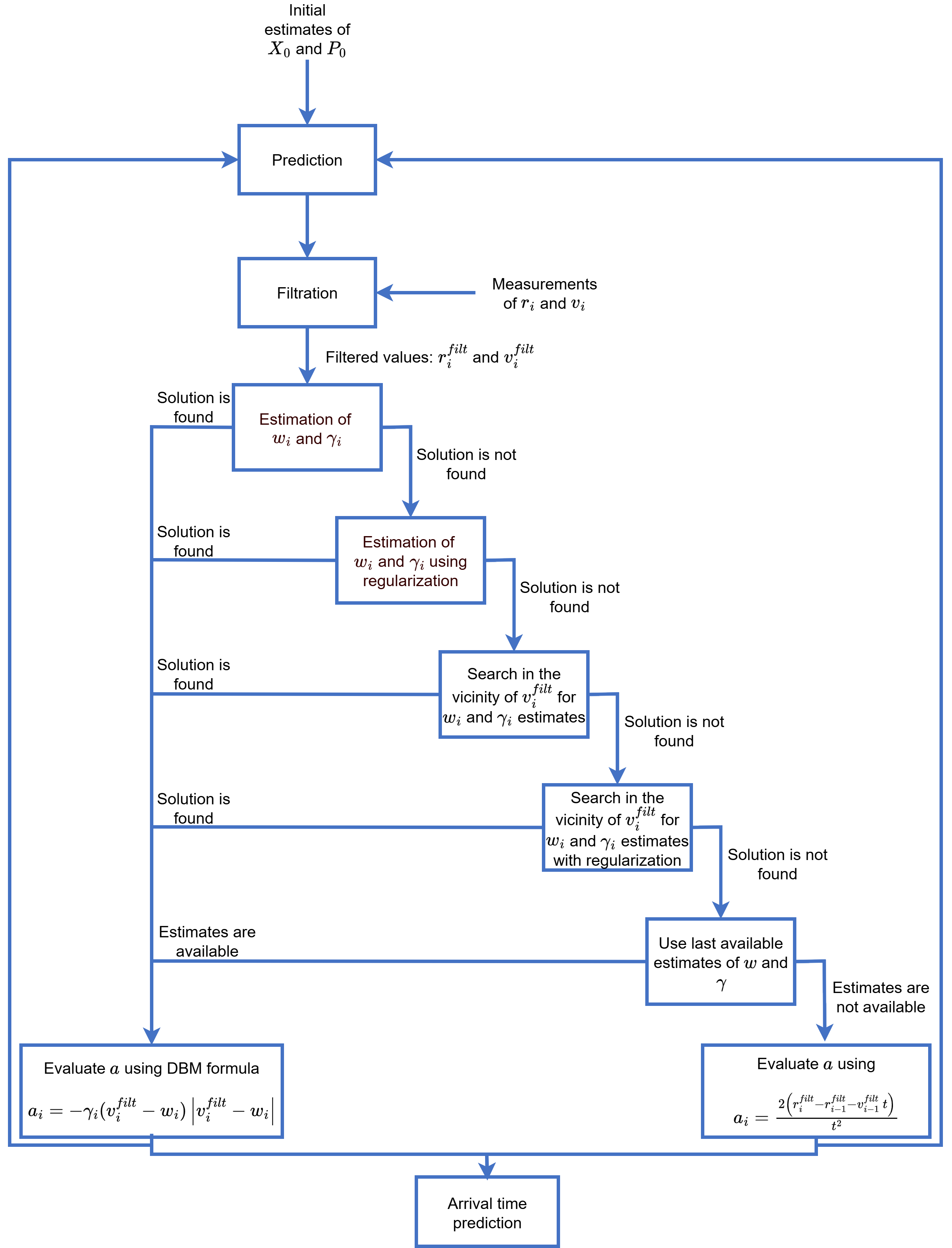}
\caption{Schematic overview of the DBM-Kalman Filter algorithm for estimating CME propagation, solar wind speed $w$ and drag parameter $\gamma$ from remote-sensing heliospheric measurements.
\label{fig:kalman_block}}
\end{figure*}

\begin{equation}\label{w_est_reg}
w = \frac{ \frac{\partial v}{\partial \gamma} (r(t) - r_0) - \frac{\partial r}{\partial \gamma} (v(t) - v_0) + \lambda_w w_{\text{prior}} }
{\frac{\partial r}{\partial w} \cdot \frac{\partial v}{\partial \gamma}-
\frac{\partial v}{\partial w} \cdot \frac{\partial r}{\partial \gamma} + \lambda_w } + w_0,
\end{equation}
\begin{equation}\label{gamma_est_reg}
\gamma = \frac{ -\frac{\partial v}{\partial w} (r(t) - r_0) + \frac{\partial r}{\partial w} (v(t) - v_0) }
{\frac{\partial r}{\partial w} \cdot \frac{\partial v}{\partial \gamma}-
\frac{\partial v}{\partial w} \cdot \frac{\partial r}{\partial \gamma} + \lambda_\gamma } + \gamma_0.
\end{equation}
Here, $\lambda_w = 1 \times 10^{-3}$, $\lambda_\gamma = 1 \times 10^{-12}$ are regularization coefficients, while $w_{prior}=500$~km/s is a prior expectation for $w$. These modifications prevent divergence and ensure stable, consistent parameter estimates even when $v_{0}$ is close to $w$.
If the estimation still fails, we perform a localized search in the vicinity of the filtered velocity $v$, defined by a range of $\pm3\sigma$, where $\sigma$ is the square root of the velocity variance component from the filtration error covariance matrix $P_{i,i}$, as described in Appendix~\ref{appendix_Kalman}.
If necessary, we repeat this search with regularization enabled. If no valid estimates of $w$ and $\gamma$ are found after all the steps, the algorithm reverts to using the most recent available estimates from earlier time steps. When we successfully determine $w$ and $\gamma$, we calculate the CME acceleration using the DBM Equation~(\ref{eq_motion_a}). If these parameters remain unavailable after all estimation steps and no valid recent estimates exist, we use an alternative expression to estimate acceleration directly from the kinematic changes in filtered position and velocity, given by the following equation:
\begin{equation}\label{acc_filter}
a_i = 2(r^{filt}_i - r^{filt}_{i-1} - v^{filt}_{i-1}t)/t^2.
\end{equation}
Finally, we obtain the predicted arrival time of the CME at the target location (e.g., Earth or Mars) by propagating the CME from time step $i$ using the most recent estimates of $w$ and $\gamma$.

To evaluate the performance of the proposed Kalman filter in estimating CME propagation dynamics, we conducted a set of six simulation scenarios, each representing a physically realistic configuration of CME propagation. These scenarios were designed to cover a range of typical conditions in interplanetary space by varying three key parameters: the initial CME speed $v_{0}$, the background solar wind speed $w$, and the drag parameter $\gamma$. Specifically, we considered three categories of CMEs based on their initial speed: slow ($v_{0} = 200$~km/s), fast ($v_{0} = 1200$~km/s), and very fast ($v_{0} = 2500$~km/s). Each CME case was tested under two different solar wind background speeds: a slow wind of $w = 300$~km/s and a faster wind of $w = 600$~km/s. The drag parameter was set to a physically appropriate value for each case; decreasing with increasing CME speed. The initial position of the CME at the start of the drag-dominated phase is chosen as $r_{0}=20R\astrosun$. 

To simulate noisy observations of CME position and velocity,
we add Gaussian noise to the true DBM-based values at each time step:
\begin{equation}\label{eq_gen_meas}
  \begin{aligned}
   r^{\text{meas}}_{i}&=r_{i}+\eta_{r} \\
   v^{\text{meas}}_{i}&=v_{i}+\eta_{v}.
  \end{aligned}
\end{equation}
Here, $r^{\text{meas}}_{i}$ and $v^{\text{meas}}_{i}$ represent the noisy measurements of the CME distance and speed at time $i$, while $r_{i}$ and $v_{i}$ are the true DBM values  (Equations~\ref{eq_DBM_r}~and~\ref{eq_DBM_v}). The standard deviations of the measurement noise, $\eta_{r}$ and $\eta_{v}$, are set to 10\% of the true values $r$ and $v$. These synthetic measurements are then assimilated into the DBM-Kalman filter described above to estimate the CME kinematics and the parameters $w$ and $\gamma$.

\begin{figure*}[!htbp]
\centering
\includegraphics[width=\textwidth]{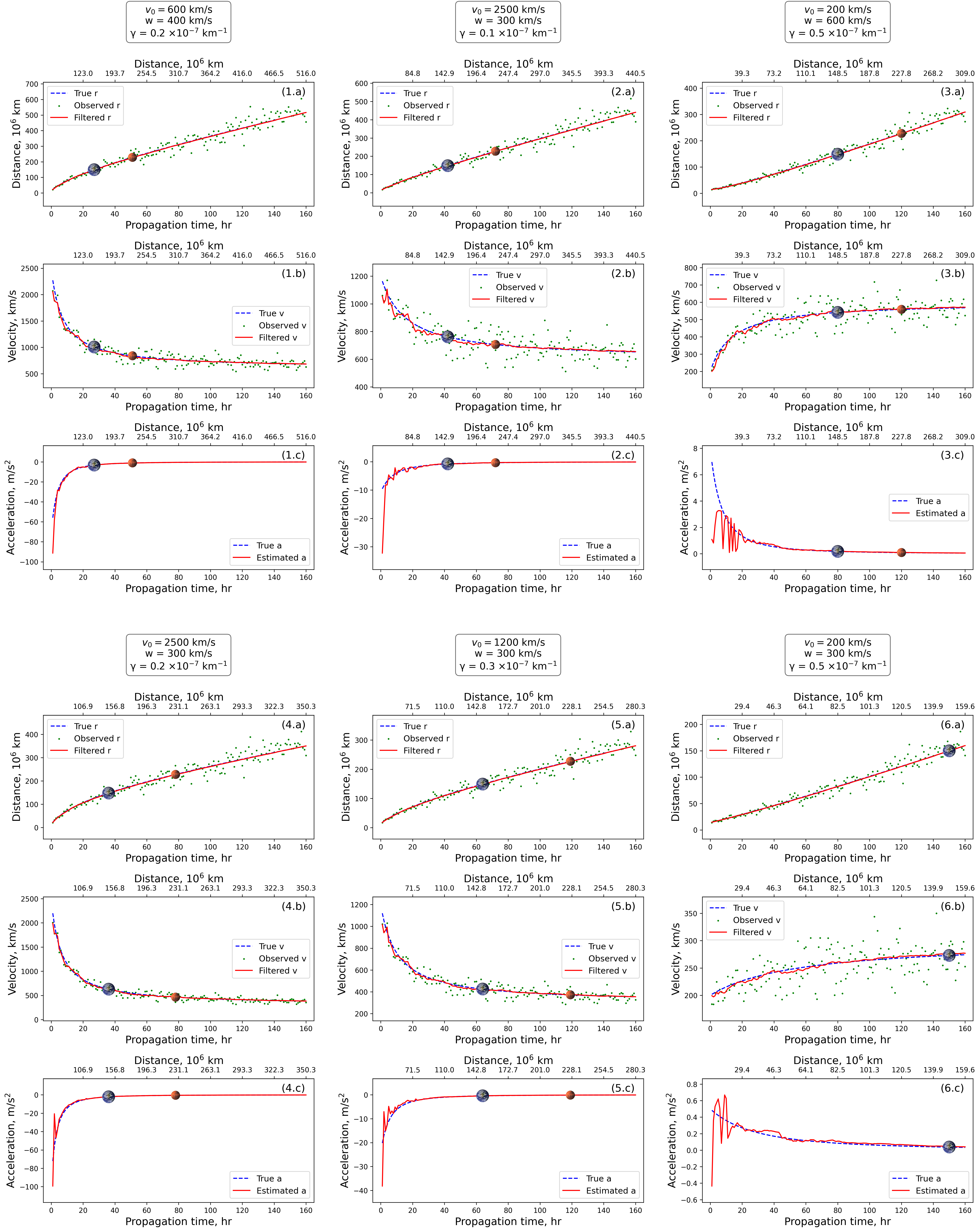}
\caption{Comparison of true (dashed blue line), observed (green dots), and filtered (solid red line) kinematic profiles for six CME propagation examples. Each group represents a different case, with three vertically stacked subplots showing the evolution of (a) heliocentric distance, (b) velocity, and (c) acceleration over time. The Earth and Mars icons indicate the arrival of the CME at the respective planets. 
\label{fig:filter}}
\end{figure*}

Figure~\ref{fig:filter} presents the results of the DBM-Kalman estimates for the CME kinematics. Each vertical block of three plots shows the CME kinematics of one of the six scenarios. In each scenario, the first row shows the true (dashed blue), observed (green dots), and filtered (solid red) CME distance. The second row displays the corresponding velocity profiles, and the third row gives the true (dashed blue) and estimated (solid red) acceleration. The true acceleration is calculated using Equation~(\ref{eq_motion_a}), and the estimated acceleration is obtained through the DBM-Kalman algorithm. Across the six scenarios, the standard deviation of the CME distance filtration errors over the full 160-hour observation period ranged from 0.23 to 0.75 million km. The velocity errors range from 3.1 to 31.3 km/s, while the acceleration errors range from 0.09 to 3.6 m/s$^2$. After 20 hours, once the rapid changes in CME acceleration have subsided and new measurements are incorporated, the errors decrease, particularly for velocity and acceleration. Distance errors drop to $0.23 - 0.73$ million km, velocity errors reduce to $2.8 - 13$ km/s, and acceleration errors reduce to 0.016 to 0.1 m/s$^2$, reflecting the improved accuracy of estimates as new measurements are incorporated. These results demonstrate that the DBM-Kalman algorithm can derive reliable estimates of CME kinematics even under high measurement noise, where distance measurements fluctuate by millions of kilometers and velocity measurements by hundreds of km/s.This demonstrates its ability to track CME propagation and predict arrival times from satellite measurements.

\begin{figure*}[!htbp]
\centering
\includegraphics[width=\textwidth]{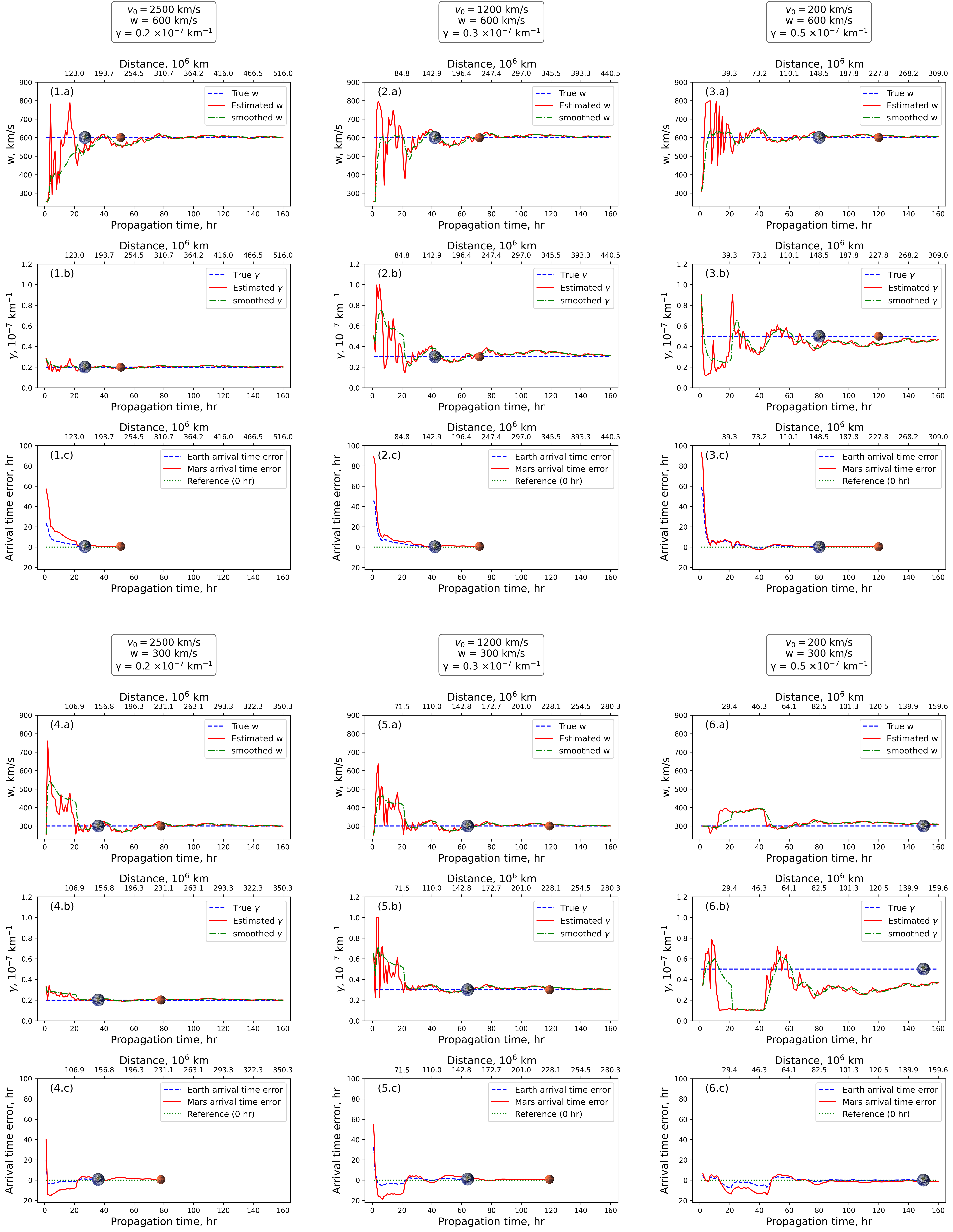}
\caption{Estimated solar wind speed $w$, drag parameter $\gamma$, and CME arrival-time error at Earth and Mars for six representative propagation cases. Each vertical group corresponds to a different case, with three stacked panels: (a) estimated (solid red) and true (dashed blue) $w$ values over time, with smoothed estimates (green dash-dot); (b) estimated (solid red), true (dashed blue), and smoothed (green dash-dot) $\gamma$ values; and (c) the arrival-time prediction error (predicted minus true). The Earth and Mars icons indicate the arrival of the CME at the respective planets. 
\label{fig:filter_time}}
\end{figure*}

Figure~\ref{fig:filter_time} presents the estimated parameters and prediction performance for the same six CME cases shown previously. Each vertical block again corresponds to one CME scenario. In each scenario, the first row shows the estimated solar wind speed $w$ (solid red) compared with its true value (dashed blue), along with a smoothed version of the estimates (green dash-dot). The second row shows the estimated drag parameter $\gamma$ (solid red) versus the true value (dashed blue), again with smoothed estimates over time (green dash-dot). The third row presents the CME arrival-time errors at each time step with blue and red curves indicating the prediction errors at Earth and Mars, respectively, computed as the difference between the predicted and the true arrival times.

Over the full 160-hour observation window, the standard deviation of $w$ errors ranges from 27.2 to 68.6~km/s, while $\gamma$ errors lie between $0.01 \times 10^{-7}$ and $0.13 \times 10^{-7}$~km$^{-1}$. However, once the initial variability in CME dynamics stabilizes after the first 20 hours, error levels decline. For $w$, the error reduces to a range of $14.2 - 27.6$~km/s, and for $\gamma$, the error narrows to between $0.007 \times 10^{-7}$ and $0.12 \times 10^{-7}$~km$^{-1}$.
At the start of the propagation, over the first $5-10$ hours when CME acceleration changes most rapidly, arrival-time errors can be relatively large as the filter adapts, reconciling model dynamics with observational data. However, the errors decrease rapidly as the event evolves. By $10-20$ hours, the strongest changes in CME acceleration have passed and arrival-time predictions improve. For example, as shown in Table~\ref{tab:error_results}, in one slow, accelerating CME case (initial speed 200 km/s; solar wind speed 600 km/s), arrival-time errors drop to $3.12 - 6.44$ hours for Earth and $2.79 - 7.5$ hours for Mars.  For very fast CMEs, the maximum deviation from the true arrival time reaches 4.94 hr for Earth ($\approx$ 18\% of the 27.13 hr transit time) and 13.03 hr for Mars ($\approx$ 26\% of the 51.03 hr transit time).

Beyond $\sim20$ hours, the errors stabilize to within 1 hour for Earth and 2 hours for Mars for most cases.
This level of accuracy is critical for mission planning, as it shows that once the early variability in CME dynamics subsides, forecasts become precise enough to guide the placement of measurement points along the CME path for effective data collection and reliable arrival-time predictions.

\begin{deluxetable*}{l c c c c c c c}
\tabletypesize{\scriptsize}
\tablewidth{0pt}
\tablecaption{Arrival-time prediction errors (in hours) at Earth and Mars across different CME initial conditions and time intervals. The table shows the minimum and maximum errors observed during three propagation phases: $5-10$ hours, $10-20$ hours, and from 20 hours until arrival. \label{tab:error_results}}
\tablehead{
\colhead{CME Parameters} & \colhead{} &
\multicolumn{2}{c}{$5 - 10$ hours} & \multicolumn{2}{c}{$10 - 20$ hours} & \multicolumn{2}{c}{20 hours $-$ arrival} \\
\colhead{} & \colhead{} &
\colhead{Earth} & \colhead{Mars} &
\colhead{Earth} & \colhead{Mars} &
\colhead{Earth} & \colhead{Mars}
}
\startdata
$v_0 = 2500$, $w_0 = 600$, $\gamma_0 = 0.2\times 10^{-7}$ & Min & 5.42 & 14.47 & 2.53 & 6.07 & 0.48 & 0.22 \\
 & Max & 7.87 & 19.93 & 4.94 & 13.03 & 2.45 & 5.95 \\
$v_0 = 1200$, $w_0 = 600$, $\gamma_0 = 0.3\times 10^{-7}$ & Min & 6.27 & 9.34 & 3.44 & 5.19 & 1.36 & 0.02 \\
 & Max & 9.19 & 14.64 & 6.24 & 10.03 & 3.63 & 5.78 \\
$v_0 = 200$, $w_0 = 600$, $\gamma_0 = 0.5\times 10^{-7}$ & Min & 2.49 & 1.91 & 3.12 & 2.79 & 0.54 & 0.05 \\
 & Max & 8.21 & 11.11 & 6.44 & 7.5 & 4.33 & 5.77 \\
$v_0 = 2500$, $w_0 = 300$, $\gamma_0 = 0.2\times 10^{-7}$ & Min & 1.72 & 9.74 & 1.3 & 8 & 0.26 & 0.02 \\
 & Max & 3.36 & 14.29 & 1.7 & 9.71 & 1.15 & 7.42 \\
$v_0 = 1200$, $w_0 = 300$, $\gamma_0 = 0.3\times 10^{-7}$ & Min & 3.22 & 13.54 & 3.39 & 13.42 & 0.52 & 0.07 \\
 & Max & 5.61 & 18.79 & 3.96 & 14.53 & 3.04 & 12.23 \\
$v_0 = 200$, $w_0 = 300$, $\gamma_0 = 0.5\times 10^{-7}$ & Min & 0.85 & 1 & 0.21 & 0.36 & 1.44 & 0.54 \\
 & Max & 3.04 & 5.31 & 7.53 & 13.24 & 7.8 & 14.44 \\
\enddata
\end{deluxetable*}

Interestingly, in the case of initial CME speed of $v_{0} = 200$~km/s, which is relatively close to the background solar wind speed $w = 300$~km/s, the estimation of the drag parameter $\gamma$ is less accurate (Figure~\ref{fig:filter_time} (6.b)) compared to cases with larger $|v_{0} - w|$. This behavior is consistent with the earlier explanation that when $V_{0} \approx w$, $r(t)$ and 
$v(t)$ are only weakly sensitive to $\gamma$, reducing its estimation accuracy. However, in this regime, the model predictions for $r$ and $v$ are only weakly sensitive to $\gamma$, as they are dominated by the solar wind speed $w$, which is reliably estimated. Consequently, the arrival-time error for this case remains small (Figure~\ref{fig:filter_time} (6.c), with a mean of 1.2 hours for Earth and 1.8 hours for Mars) despite the reduced accuracy in $\gamma$. In contrast, when the difference between $v_{0}$ and $w$ is large, the drag parameter $\gamma$ is estimated more reliably, which is particularly important for fast CMEs, as they pose the greatest risks for space weather in the heliosphere.

These results demonstrate that the DBM-Kalman algorithm, even under such noisy measurements, enables a meaningful reconstruction of CME kinematics and reliable arrival-time forecasts for interplanetary targets, provided that the CME kinematics are fully captured by the drag-based model.

\section{Conclusions}
We developed HELIOPANDA, a next-generation CME arrival forecasting framework that integrates the DBM with spacecraft observations through a nonlinear assimilation scheme. The framework advances in three steps. First, we validated the method for estimating the critical parameters governing CME propagation, the solar wind speed $w$ and drag parameter $\gamma$, directly from pairs of distance and speed observations, then applied it to in-situ probe scenarios, and finally extended it to heliospheric remote-sensing data assimilation.

Our simulations, covering 4,480 synthetic CME profiles, demonstrated that the analytical structure of DBM enables reliable recovery of $w$ and $\gamma$, providing stable and highly accurate estimates across various CME and solar wind conditions for all test cases, errors remained below $0.0002~\mathrm{km/s}$ for $w$ and $0.000004 \times 10^{-7}$~km$^{-1}$ for $\gamma$, confirming its robustness and providing a basis for application to in-situ and noisy remote-sensing spacecraft observations.

Building on these results, we examined the value of in-situ measurements by introducing nine virtual spacecraft along the Sun–Earth line, each treated independently. We showed that CME arrival predictions at Earth can achieve errors as low as 0.6 hours for moderate-speed CMEs (600 km/s) and 1 hour for fast CMEs (2500 km/s) when the spacecraft is located 30 million km from the Sun, with accuracy improving as the CME is tracked farther out. Even a single probe at intermediate heliocentric distances delivers early predictions with substantial accuracy, supporting early warning capabilities. This controlled setup highlights the potential of missions that provide distributed sampling of CME kinematics.

Extending the framework further, we assimilated heliospheric remote-sensing measurements, which unlike in-situ data are inherently noisy and require additional treatment. By assimilating 160 noisy heliospheric measurements taken at a cadence of one hour with a Kalman filter, HELIOPANDA dynamically reconstructed CME kinematics and DBM parameters, achieving arrival-time predictions at Earth and Mars with errors well within 2 hours for both moderate and fast CME cases. The combination of analytical modeling, iterative parameter estimation, and data assimilation bridges the gap between empirical and MHD-based forecasting approaches.

Together, these steps define the HELIOPANDA framework for CME arrival forecasting  progressing from validation of the DBM solver, through the benefits of distributed in-situ probes, to applications with noisy remote-sensing data. This progression bridges the gap between empirical approaches and full MHD simulations, providing a foundation for real-time, multi-point CME forecasting with the current and future heliophysics missions. The proposed framework establishes the basis for continuous monitoring of solar activity that can leverage the expanding constellation of heliophysics missions, with integration into operational space weather services promising more reliable and timely alerts for planetary environments, safeguarding satellites, power grids, and human and robotic space exploration.

Future research should explore the model’s performance under conditions where only CME distance measurements are available, as well as under varying observational cadences. Tests with real CME events and assimilation of in-situ and remote-sensing measurements from ongoing missions, including coronagraphs and heliospheric imagers aboard STEREO, Solar Orbiter, Parker Solar Probe, PUNCH, and upcoming L4/L5 platforms will be essential to further refine and operationalize the method.

\begin{acknowledgments}
This work was supported by internal center funding. A.M.V. and M.D. acknowledge the support from the Austrian-Croatian Bilateral Scientific Project ``Analysis of solar eruptive phenomena from cradle to grave''. We thank the reviewers for valuable comments on this study.
\end{acknowledgments}

\begin{contribution}
Z.A. and T.P. developed the method and led the writing of the paper. A.M.V., M.D., and S.J.H. contributed to conceptualization, study development, analysis, and writing. All authors discussed the results and provided feedback on the manuscript.
\end{contribution}

\appendix

\section{The Kalman filter algorithm}\label{appendix_Kalman}
The Kalman filter is a recurrent algorithm that operates in two steps, incorporating the following procedures:

1. \textit{Prediction} is performed to estimate the future state vector $X$ one step ahead:
\begin{equation}\label{eq_extrapolation}
X_{i,i-1}=\Phi_{i,i-1}X_{i-1,i-1}+Ga_{i-1}.
\end{equation}
Here, $X_{i,i-1}$ denotes the predicted estimate of state $X_{i}$ at step $i$. The first subscript $i$ marks the time of estimating the state vector $X_{i}$, while the second subscript $i-1$
specifies the number of $r^{meas}$ and $v^{meas}$ observations 
assimilated by the algorithm for the prediction. Similarly, $X_{i-1,i-1}$ represents the filtered estimate of state $X_{i-1}$ at the previous step $i-1$, where the second subscript $i-1$
indicates that this filtered estimate is derived using the measurements up to the time step $i-1$. 

The prediction accuracy of $X_{i,i-1}$ is quantified through the prediction error covariance matrix:
\begin{equation}\label{eq_P_extr}
P_{i,i-1}=\Phi_{i,i-1}P_{i-1,i-1}\Phi_{i,i-1}^T+Q_{i}.
\end{equation}
Here, diagonal elements of matrix $P_{i,i-1}$ represent the variances of the predicted state components, while off-diagonal elements indicate their mutual correlations. The matrix $Q_{i}$ denotes the covariance matrix of the state noise 
$\epsilon$ in Equation~(\ref{state_eq}), reflects model uncertainties in CME propagation dynamics with diagonal elements representing variances of the state noise for each component of state vector. At the first prediction step, $Q$ is initialized as $Q=G\sigma_a^2G^T$, where $\sigma_a$ is set to $1$ for fast CMEs (to favor measurements due to rapid dynamical changes) and $0$ for slow CMEs, where trusting the model is preferable.

2. Filtration is performed to incorporate new measurements $z=[r^{mes}, v^{mes}]^T$ , yielding an improved estimate of the state $X_{i}$ at step $i$:
\begin{equation}\label{eq_P_extr}
X_{i,i-1}=X_{i,i-1}+K_{i}(z_{i}-HX_{i,i-1}).
\end{equation}
Here, $K_{i}$ is the filter gain that determines the relative weight assigned to the current measurement $z_{i}$, and the predicted state $X_{i,i-1}$:
\begin{equation}\label{eq_K_gain}
K_{i}=P_{i,i-1}H_{i}^T(H_{i}P_{i,i-1}H_{i}^T+R_{i})^{-1}.
\end{equation}
Here, $R_{i}$ describes the covariance matrix of the measurement noise $\eta_{r}$ and $\eta_{v}$ in Equation~(\ref{meas_eq}). The accuracy of the filtered state $X_{i,i}$ is characterized by the filtration error covariance matrix:
\begin{equation}\label{eq_P_extr}
P_{i,i}=(I-K_{i}H_{i})P_{i,i-1}.
\end{equation}
Here, $I$ is the $2\times2$ identity matrix. 

The Kalman filter offers optimal estimation by minimizing the covariance of the estimation error, regardless of the noise distribution \citep{Kalman1960,Sage1971}. However, accurately specifying the model error statistics is essential, as uncertainties in the model or measurements can lead to significant prediction errors and potentially cause the Kalman filter to fail. In practice, these noise statistics can be derived from dedicated statistical studies or identified directly from measurements using specialized methods \citep{Podladchikova2012_Kalman_Sunspot,
Podladchikova_2014_RB1,Podladchikova2014_RB2,Petrova2021}.

In this study, the standard deviations of the measurement noise $\eta_r$ and $\eta_v$ are set to 10\% of either reference or filtered values of $r$ and $v$ as a working assumption, though they can be refined using the above identification approaches. 
The state noise covariance matrix $Q_{i}$ is updated adaptively at each assimilation step based on the estimated solar wind speed $w_{i}$ and drag parameter $\gamma_{i}$. The acceleration uncertainty $\sigma_a^2$ is derived using a first-order Taylor expansion of Equation~\ref{eq_motion_a} around the current estimates of $w$ and $\gamma$. Linearizing $a$ with respect with these parameters gives the following equation:
\begin{equation}\label{eq_sigma_a}
\sigma_a^2 =
\left(\frac{\partial a}{\partial \gamma}\right)^2 \sigma_\gamma^2 +
\left(\frac{\partial a}{\partial w}\right)^2 \sigma_w^2.
\end{equation}
Here, $\frac{\partial a}{\partial \gamma}=-(v_i-w_i)|v_i-w_i|$, $\frac{\partial a}{\partial w}=2\gamma_i|v_i-w_i|$, and $\sigma_w^2$ and $\sigma_\gamma^2$ are variances of estimation errors for $w$ and $\gamma$. The resulting $\sigma_a^{2}$ defines $Q_i$ as 
\begin{equation}\label{eq_Q_update}
Q_i = \sigma_a^2
\begin{bmatrix}
t^4/4 & t^3/2 \\
t^3/2 & t^2
\end{bmatrix}.
\end{equation}
Here, $t$ is the time step. This adaptive update allows the filter to account for temporal changes in dynamical uncertainty, preventing divergence and ensuring that the Kalman gain $K_{i}$ adjusts between prediction and measurement.

\begin{equation}\label{eq_sigma_w_gamma_spin_up}
\sigma_w = \alpha\,w_i, \quad \sigma_\gamma = \alpha\,\gamma_i.
\end{equation}
Here, $\alpha=0.5$ when the solution is obtained via the iterative procedure described in Section~\ref{Section_w_gamma_est}, and $\alpha=0.7$ when it is derived from a search in the vicinity of the filtered velocity. If the CME speed is close to the estimated solar wind speed $|v_0-w_i|\leq100~$~km/s, the process noise is reduced to avoid overestimating the uncertainty. For particularly slow CMEs ($v_{0}=200$~km/s under this condition), the uncertainties are further reduced to $\alpha=0.1$ to stabilize the filter.

Once the initial transient phase is over (after around 20 hours of observations), $\sigma_w$ and $\sigma_\gamma$ are no longer set by fixed percentages, but instead are dynamically estimated as the running standard deviations of all prior parameter estimates:
\begin{equation}\label{eq_sigma_w_gamma_later}
\sigma_w = \sqrt{\mathrm{var}(w_{1:k})}, \qquad
\sigma_\gamma = \sqrt{\mathrm{var}(\gamma_{1:k})}.
\end{equation} 

This approach allows the process noise covariance matrix to reflect the accumulated statistical variability of $w_{i}$ and $\gamma_{i}$, enabling $Q_{i}$ to gradually adapt to the actual uncertainty in the system as more observations become available. These adaptive rules prevent overconfidence during the early high-acceleration stage and ensure that the filter maintains physically consistent uncertainty estimates throughout the CME evolution.

\bibliography{My_References}{}
\bibliographystyle{aasjournalv7}

\end{document}